\documentclass{JHEP3}

\usepackage{graphicx,times}
\usepackage{multirow}
\usepackage{amsmath,amssymb}
\usepackage{epsfig,multicol,bbm}

\newcommand{\ub}{\overline{u}}
\newcommand{\db}{\overline{d}}

\title{Anomaly and a QCD-like phase diagram \\ 
with massive bosonic baryons}
\author{Shailesh Chandrasekharan and Anyi Li\\
  Department of Physics, Box 90305, Duke University,
  Durham, North Carolina 27708, USA
  E-mail: \email{sch@phy.duke.edu,anyili@phy.duke.edu}}

\abstract
{
We study a strongly coupled $Z_2$ lattice gauge theory with two flavors of quarks, invariant under an exact $\mathrm{SU}(2)\times \mathrm{SU}(2) \times \mathrm{U}_A(1) \times \mathrm{U}_B(1)$ symmetry which is the same as QCD with two flavors of quarks without an anomaly. The model also contains a coupling that can be used to break the $\mathrm{U}_A(1)$ symmetry and thus mimic the QCD anomaly. At low temperatures $T$ and small baryon chemical potential $\mu_B$ the model contains massless pions and massive bosonic baryons similar to QCD with an even number of colors. In this work we study the $T-\mu_B$ phase diagram of the model and show that it contains three phases : (1) A chirally broken phase at low $T$ and $\mu_B$, (2) a chirally symmetric baryon superfluid phase at low $T$ and high $\mu_B$, and (3) a symmetric phase at high $T$. We find that the nature of the finite temperature chiral phase transition and in particular the location of the tricritical point that seperates the first order line from the second order line is affected significantly by the anomaly.}


\begin{document}

\section{Introduction}

Understanding the QCD phase diagram in the temperature $T$ and baryon chemical potential $\mu_B$ plane is important from both phenomenological and experimental point of view \cite{RevModPhys.80.1455,Munzinger07,Aggarwal:2010cw}. Based on very general arguments it is possible to predict the important phases and the rough structure of the phase diagram \cite{Halasz:1998qr,Berges:1998rc}. However, for a quantitative understanding of the phase diagram and to determine the location of critical points, careful non-perturbative calculations within the framework of QCD are necessary. While the finite temperature phase transition at $\mu_B=0$ can be understood well \cite{Bazavov:2009zn,Aoki:2006we,Aoki:2006br,Aoki:2009sc,Borsanyi:2010bp}, a reliable first principles calculation at $\mu_B\neq 0$ is impossible today due to the notorious sign problem. Despite this difficulty, many exploratory studies of the phase diagram have emerged over the past decade \cite{Philipsen:2008gf,deForcrand:2010ys,Lombardo:2009tf,Gavai:2008zr}. Since the systematic errors in all these studies are far from control, it is fair to say that important questions about the QCD phase diagram remain unanswered. For example, where is the critical point \cite{Stephanov:2004wx}? Can nuclear matter form a crystalline state \cite{PhysRevD.63.074016}? Can there be a state of matter where chiral symmetry is restored while the quarks remain confined (a quarkyonic phase) \cite{McLerran:2007qj}? Until a solution to the QCD sign problem is found these questions may not be answered satisfactorily.

Given the difficulty of uncovering the QCD phase diagram from first principles, many studies have focused on model studies. These include random matrix models \cite{Jackson:1995nf,Sano:2009wd,Stephanov:2006dn}, NJL models \cite{Hands:2002mr,Buballa:2003qv,Sun:2007fc,Fukushima:2008wg}, PNJL and related models \cite{Ratti:2005jh,Schaefer:2007pw,Mao:2009aq,Cristoforetti:2009tb}, sigma models \cite{Schaefer:2008hk,Bowman:2008kc} and strong coupling lattice models \cite{Nishida:2003uj,Kawamoto:2005mq}. Some have also focused on two color lattice QCD \cite{Kogut:1999iv,Kogut:2001if,Kogut:2001na,Kogut:2002cm,Chandrasekharan:2006tz,Andersen:2010vu} since it does not suffer from the sign problem and still has an interesting phase diagram \cite{Kogut:2000ek}. Effects of the anomaly on the phase diagram have also been considered in model calculations \cite{Brauner:2009gu,Chen:2009gv,Hatsuda:2006ps,Yamamoto:2007ah}. While model studies cannot give quantitative information about the QCD phase diagram, they are still useful to understand the nature of possible phases and critical points. Unfortunately, if one wishes to study these models from first principles using Monte Carlo methods, one again encounters sign problems. Hence most studies so far involve the mean field approximation. It would be useful to go beyond this approximation at least within models calculations. Recently, in a pioneering study, the $T-\mu_B$ phase diagram of the strong coupling limit of lattice QCD with staggered fermions was computed \cite{deForcrand:2009dh}. Some of the features of how nuclei emerge in lattice QCD was illustrated. Although the nuclear binding energy was found to be large compared to the real world, such ab initio calculations of nuclear matter from model field theories could teach us valuable lessons.

In this work we continue this trend to perform ab initio calculations in model field theories and study a toy model of QCD with an even number of colors that contains massless pions and massive bosonic baryons. The $T-\mu_B$ phase diagram of QCD with an even number of colors is also expected to be complex and interesting. When the baryon chemical potential exceeds a critical value, baryons will condense to form a superfluid state. If this phase transition is first order, then the corresponding transition line can end on a second order critical point in the $T-\mu_B$ plane just like in real QCD \cite{Stephanov:2004wx}. QCD with an even number of colors will also have exotic phases such as the quarkyonic phase and color superconductivity. When the number of colors is greater than two, baryons are naturally massive while the pions are massless. Two color QCD is special since baryons and pions are both massless due to extra symmetries that transform baryons into pions \cite{Kogut:1999iv}. By breaking these symmetries baryons can be made massive while keeping pions massless in a lattice field theory. As far as we know, field theory models with massless pions and massive bosonic baryons have not been studied in the literature with Monte methods since they too suffer from sign problems in the conventional approach.

World line formulations of lattice field theory offer new hope since they lead to new solutions to the sign problem \cite{Chandrasekharan:2008gp} and allow us to accurately compute phase diagrams of lattice field theories with new algorithms \cite{Prokof'ev:2001zz,Syljuasen:2002zz,Adams:2003cc,Wolff:2010zu}. Recently, a model of two flavor QCD, invariant under $SU(2) \times SU(2) \times U_A(1) \times U_B(1)$, was constructed in the world line formulation\cite{Chandrasekharan:2007up,PhysRevD.77.014506}. The model contains a coupling that can be used to break the anomalous $U_A(1)$ symmetry and thus mimic the QCD anomaly. In this work we extend the model to include massive baryons. Since there is no sign problem we can study the $T-\mu_B$ phase diagram accurately. We find that our model indeed contains a QCD-like tricritical point but its location strongly depends on the strength of the anomaly. We also observe that the model contains a confined but chirally symmetric baryon superfluid phase similar to the quarkyonic phase at high $\mu_B$ and low $T$.

Our paper is organized as follows. In section \ref{model} we describe the model in detail along with its world line representation. We also discuss observables that we use to identify and distinguish the phases in the model. In section \ref{zeroanomaly} we discuss our results of the phase diagram in the absence of the anomaly. In section \ref{largeanomaly} we discuss how the phase diagram changes in the presence of a large anomaly and how this change comes about as the anomaly is increased slowly. Section \ref{concl} contains our conclusions.

\section{Model and Observables}
\label{model}

The model we study is a $Z_2$ lattice gauge theory on a four dimensional hypercubic lattice whose action is given by
\begin{eqnarray}
S &=& 
-  \sum_{x,\alpha} \ t_\alpha\ 
\Bigg[ 
(\overline{\psi}_x \psi_{x+\alpha})(\overline{\psi}_{x+\alpha} \psi_x)
+ \frac{\mathrm{e}^{-m_B}}{2} \Big\{
\mathrm{e}^{\mu_B \delta_{\alpha,t}}(\overline{\psi}_x \psi_{x+\alpha})^2
+ \mathrm{e}^{-\mu_B \delta_{\alpha,t}}(\overline{\psi}_{x+\alpha} \psi_x)^2
\Big\}\Bigg]
\nonumber \\
& & 
\hskip1.0in \ + \delta \ \sum_{x,\alpha} \frac{(t_\alpha)^2}{2}
\Big\{(\overline{\psi}_x \psi_{x+\alpha})
(\overline{\psi}_{x+\alpha} \psi_x)\Big\}^2\ 
\ -\  \frac{c}{2} \sum_x (\overline{\psi}_x\psi_x)^2 
\label{action}
\end{eqnarray}
where $x$ denotes a lattice site on an $L^3 \times L_t$ hypercubic lattice and $\alpha=1,2,3,4$ denotes the direction. The fields $\psi_x$ and $\overline{\psi}_x$ are two component Grassmann valued fields given by
\begin{equation}
\psi_x = 
\left(\begin{array}{c} 
u_x \cr d_x \end{array}\right),\ \ 
\overline{\psi}_x
=  \left(\begin{array}{cc} \overline{u}_x & \overline{d}_x
\end{array}\right)
\end{equation}
We choose $t_\alpha = 1$ for $\alpha = 1,2,3$ and $t_4 = T$. The asymmetry factor $T$ controls the temperature, the coupling $c$ controls the strength of the anomaly, the parameter $m_B$ controls the mass of the baryon and the parameter $\mu_B$ is the baryon chemical potential. The parameter $\delta$ is fixed based on convenience as discussed later.

The most interesting feature of the action (\ref{action}) is that it shares the global symmetries of two flavor QCD.  When $c=0$ the action is invariant under a global $SU(2)\times SU(2)\times U_B(1) \times U_A(B)$ symmetry.  Indeed it is easy to check that the action is invariant under the transformation
\begin{subequations}
\begin{eqnarray}
\psi_{x_e} \rightarrow \mathrm{e}^{i\theta_A+i\theta_B}L\psi_{x_e},\ \ \ 
&& \psi_{x_o} \rightarrow \mathrm{e}^{-i\theta_A+i\theta_B}R\psi_{x_o},
\\
\overline{\psi}_{x_o} \rightarrow \overline{\psi}_{x_o} L^\dagger\mathrm{e}^{-i\theta_A-i\theta_B},\ \ \ 
&&\overline{\psi}_{x_e} \rightarrow \overline{\psi}_{x_e} R^\dagger \mathrm{e}^{i\theta_A-i\theta_B},
\end{eqnarray}
\end{subequations}
where $L,R\in SU(2)$ and $x_e$ and $x_o$ refer to even and odd sites. When $c\neq 0$, the $U_A(1)$ symmetry is explicitly broken and the action is invariant only under $SU(2) \times SU(2) \times U_B(1)$. We interpret the $U_A(1)$ symmetry as the anomalous axial symmetry of QCD and the parameter $c$ will be used to change its strength of the anomaly \cite{Chandrasekharan:2007up,PhysRevD.77.014506}. The $U_B(1)$ is the baryon number symmetry and the chemical potential $\mu_B$ couples to the baryon numbers as in QCD. In addition to the global symmetries the action is invariant under a $Z_2$ gauge transformation $\psi_x \rightarrow - \psi_x$ and $\overline{\psi}_x \rightarrow - \overline{\psi}_x$ at every site $x$. Thus, the baryons of the model are bosons made up of confined $ud$-diquarks. 

\FIGURE[t]{
\includegraphics[width=0.8\textwidth]{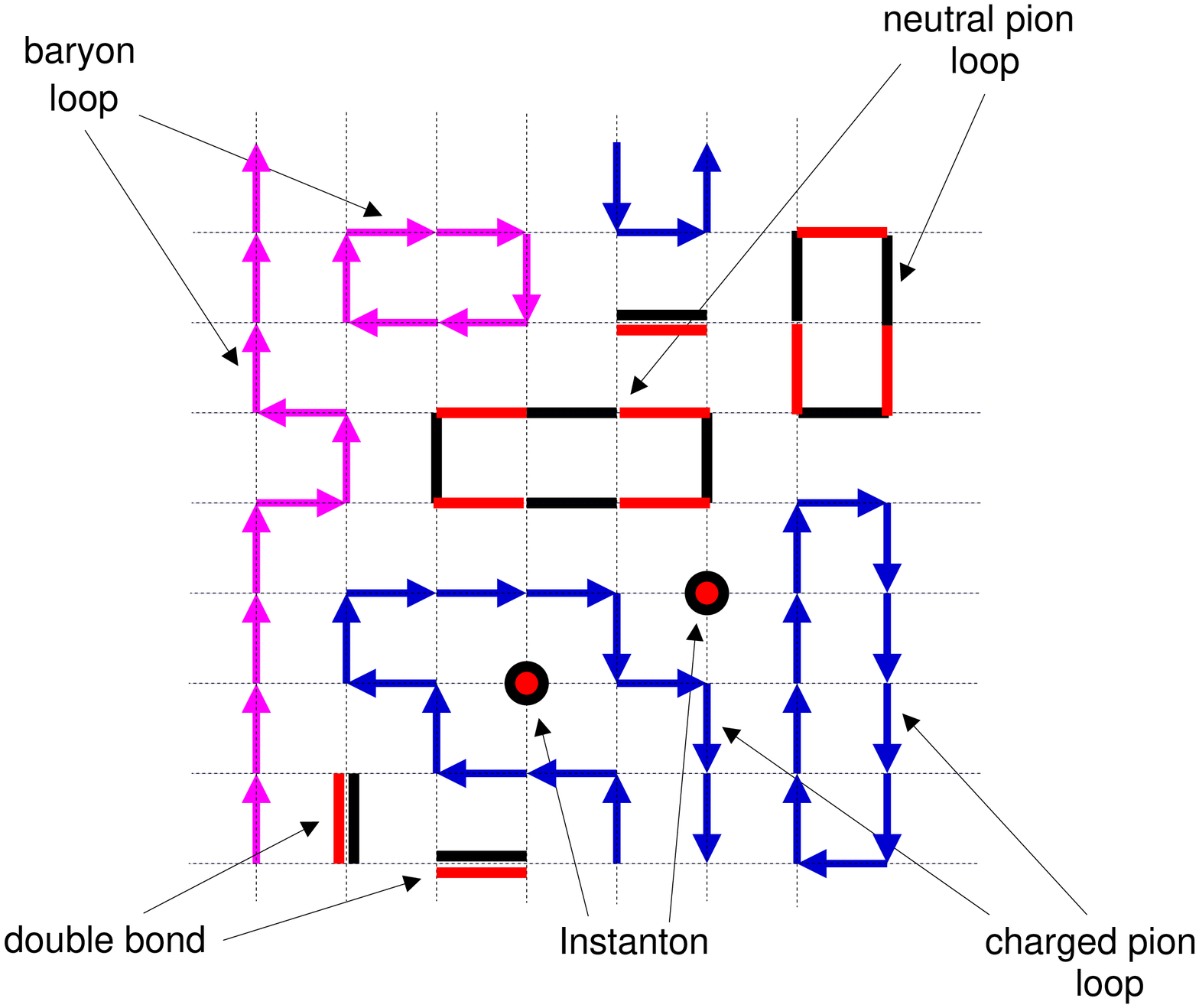}
\caption{An example of the world line configuration of pions and baryons. The five types of fermion bags, as discussed in the text, are also shown.\label{fig1}}
}

After integrating over the Grassmann variables, the partition function of the model can be written as a sum over world line configurations of neutral pions ($\ub u, \db d$), charged pions ($\ub d, \db u)$ and baryons ($u d, \db \ub$). Each configurations can contain five types of objects or fermion bags \cite{Chandrasekharan:2009wc,Chandrasekharan:2010iy}. These are (1) instantons (isolated sites), (2) double-bonds (isolated bonds connecting neighboring sites), (3) neutral-pion loops (4) charged-pion loops, and (5) baryon loops. The loops are self-avoiding loops since they represent hard-core bosons. An example of a world line configuration with all the five types of fermion bags is shown in figure~\ref{fig1}. The weight of each configuration is the product of the weights of all the fermion bags in the configuration. Instantons have a weight $c$, double-bonds have a weight $2 (t_\alpha)^2 (1+\mathrm{e}^{-2 m_B}/2-\delta)$, and all the loops have a weight $T^{n_t}$ where $n_t$ stands for the number of time-like bonds in the bag. The baryon loops are weighted with an additional factor $\exp(-m_B n - \mu_B w)$ where $n$ is the total number of bonds in the loop and $w$ is the temporal winding of the loop. We will define $\delta = (\mathrm{e}^{-2 m_B}+2-\omega_D)/2$, so that the double-bonds have a weight $\omega_D(t_\alpha)^2$.

A variety of observables can be measured with our algorithm. However, for simplicity, we will mainly focus on the following three current-current susceptibilities:
\begin{enumerate}
\item Chiral current susceptibility
\begin{equation}
Y_C = \frac{1}{L^3}\sum_{x,y}\langle J^C_\alpha(x) J^C_\alpha(y)\rangle
\end{equation}
\item Axial current susceptibility
\begin{equation}
Y_A = \frac{1}{L^3}\sum_{x,y}\langle J^A_\alpha(x) J^A_\alpha(y)\rangle
\end{equation}
\item Baryon current susceptibility
\begin{equation}
Y_B = \frac{1}{L^3}\sum_{x,y}\langle J^B_\alpha(x) J^B_\alpha(y)\rangle.
\end{equation}
\end{enumerate}
where $J^C_\alpha(x)$ is the conserved chiral current, $J^A_\alpha(x)$ is the conserved axial current and $J^B_\alpha(x)$ is the conserved baryon current. Note that these currents are variables that are defined on the bonds connecting the site $x$ with $x+\hat{\alpha}$. These currents can be measured easily in the world line configuration and Ref.~\cite{PhysRevD.77.014506} contains a discussion of how $J^C_\alpha$ and $J^A_\alpha$ are measured. $J^B_\alpha$ can similarly be measured by following the worldline of baryons. For example in figure \ref{fig1}, $J^B_\alpha(x)$ is zero on bonds that do not contain a baryon loop and on bonds which contain the baryon loop it takes value $+1$ or $-1$ depending on the direction of the baryon loop. The time component of the baryon current is the baryon number and its average over the lattice gives the baryon density $\rho_B$.

Using the current susceptibilities $Y_i$ defined above, it is easy to determine if the symmetry generated by the specific current $J^i,i \equiv C,A,B$ is spontaneously broken or not. One expects
\begin{equation}
\lim_{L\rightarrow\infty} Y_i= \left \lbrace
\begin{array}{c c}
  \rho_i \neq 0 &\text{Broken phase}\\
  A\exp(-aL) & \text{Symmetric phase}
\end{array}
\right.
\label{eq:large_behavior}
\end{equation}
In our analysis we will use this finite size scaling behavior to distinguish the phases. We will also need to distinguish between a first order and a second order phase transition between the various phases. This is accomplished by looking for the critical scaling relation
\begin{equation}
 L Y_i(L,T) \sim \sum_k ~f_k\ [(T_c-T)L^{1/\nu}]^k.
 \label{eq:scaling}
\end{equation}
which must hold close to a second order critical point $T_c$. In other words, if we plot $LY_i$ as a function of $T$ for different values of $L$ we expect all the lines to cross at a single point at a second order critical at $T=T_c$ for large values of $L$. If this does not occur we claim the transition to be first order. In addition we also look for two state signals in the observables close $T_c$ to confirm the first order nature of the transition.

We update the world line configurations using the directed path algorithm discussed in \cite{Adams:2003cc,PhysRevD.77.014506}. We have tested the algorithm against exact results on a $2 \times 2$ lattice and the comparison is given in the appendix. Below we present results obtained on an $L^3 \times 4$ lattice. We fix $m_B=0.1$ and $\omega_D = 1.0$ throughout the study for convenience. For these parameters we find that the renormalized baryon mass $M_B$ obtained through a baryon-baryon correlation function depends on $T$ but is bounded in the range $0.8 \geq M_B \geq 0.5$.

\section{Phase Diagram without Anomaly}
\label{zeroanomaly}

We begin with $c=0$ so that the action is invariant under the $U_A(1)$ symmetry and the anomaly is absent. This is similar to QCD with an infinite number of colors. At small $T$ and $\mu_B$ the $SU(2)\times SU(2) \times U_A(1)$ symmetry is spontaneous broken into the diagonal $SU(2)$ flavor symmetry leading to four Goldstone bosons. Since the baryons are massive the vacuum is free of them and the $U_B(1)$ symmetry remains unbroken. Thus, as $L$ increases, based on Eq.~(\ref{eq:large_behavior}) we expect $Y_C$ and $Y_A$ will go to non-zero constants while $Y_B$ will go to zero exponentially. At high temperatures, when all symmetries are restored, all three susceptibilities should go to zero exponentially. These expectations are consistent with our findings (see figure \ref{fig2} where the $\mu_B=0$ results are plotted). 

\FIGURE[b]{
\begin{tabular}{cc}
\begin{minipage}[c]{0.45\textwidth}
\includegraphics[width=\textwidth]{fig2a}
\end{minipage}
&
\begin{minipage}[c]{0.45\textwidth}
\includegraphics[width=\textwidth]{fig2b}
\end{minipage}
\end{tabular}
\caption{Plot of $Y_i,i=A,B,C$ as a function of $L$ at $c,\mu_B=0$ and $T=2.16$ (left) and  $T=2.19$ (right). \label{fig2}.}
}

\FIGURE[t]{
\begin{tabular}{cc}
\begin{minipage}[c]{0.45\textwidth}
\includegraphics[width=\textwidth]{fig3a}
\end{minipage}
&
\begin{minipage}[c]{0.45\textwidth}
\includegraphics[width=\textwidth]{fig3b}
\end{minipage}
\\
\begin{minipage}[c]{0.45\textwidth}
\includegraphics[width=\textwidth]{fig3c}
\end{minipage}
&
\begin{minipage}[c]{0.45\textwidth}
\includegraphics[width=\textwidth]{fig3d}
\end{minipage}
\\
\begin{minipage}[c]{0.45\textwidth}
\includegraphics[width=\textwidth]{fig3e}
\end{minipage}
&
\begin{minipage}[c]{0.45\textwidth}
\includegraphics[width=\textwidth]{fig3f}
\end{minipage}
\end{tabular}
\caption{Plots of $LY_C$ (left) and $LY_A$ (right) as a function of $T$ at different values of $L$. The data shown is for $c=0.0$ and $\mu_B=0.0$ (top row) $\mu_B=0.2$ (middle row) and $\mu_B=0.4$ bottom row.\label{fig3}}
}

In order to identify the nature of the chiral phase transition we check if Eq.~(\ref{eq:scaling}) holds. In figure~\ref{fig3}, we plot $LY_{A,C}(L)$ for $L=12, 16, 24, 32$ as function of $T$ at $\mu_B=0$, $0.2$ and $0.4$. Although there is a clear indication for a phase transition, it not consistent with second order scaling. In particular we do not see the lines at different values of $L$ cross at a single point. If the transition was first order we also expect a two state signal at the transition point. In figure~\ref{fig4} we plot $Y_{A}$ and $Y_C$ at $\mu_B=0.0$ as function of Monte Carlo time at $T=2.168$ and $L=32$. The two state signal is clearly visible confirming that the transition is indeed first order with $T_c \approx 2.168$ at $\mu_B=0.0$.

\FIGURE[t]{
\includegraphics[scale=0.4]{fig4}
\caption{Monte Carlo time evolution of $Y_C$ and $Y_A$ at $L = 32, T = 2.168$ and $c,\mu_B=0$. Tunneling between two meta-stable states which is characteristic of a first order phase transition is clearly seen.\label{fig4}}
}

For a fixed but small $T$, as $\mu_B$ increases we find that the baryon density $\rho_B$ jumps at a critical value of $\mu_B$ (which depends weakly on $T$). At the same critical value $Y_B$ also jumps to a non-zero value while $Y_c$ drops to zero. On the other hand $Y_A$ remains non-zero on both sides but also shows a small jump. As an illustration of these results, we show the data at $T=1.6$ in figure \ref{fig5}. These results confirm that at large $\mu_B$ and small $T$ the $SU(2)\times SU(2)$ symmetry is restored while $U_B(1)$ and $U_A(1)$ are both spontaneously broken. In other words the high $\mu_B$ phase is a baryon superfluid. Since baryons in our model carry axial charge, it is not surprising that $U_A(1)$ is also spontaneously broken along with $U_B(1)$ in the superfluid phase.

\FIGURE[t]{
\begin{tabular}{cc}
\begin{minipage}[c]{0.45\textwidth}
\includegraphics[width=\textwidth]{fig5a}
\end{minipage}
&
\begin{minipage}[c]{0.45\textwidth}
\includegraphics[width=\textwidth]{fig5b}
\end{minipage}
\\
&
\\
\begin{minipage}[c]{0.45\textwidth}
\includegraphics[width=\textwidth]{fig5c}
\end{minipage}
&
\begin{minipage}[c]{0.45\textwidth}
\includegraphics[width=\textwidth]{fig5d}
\end{minipage}
\end{tabular}
\caption{Plot of the $Y_C$ (top left), $Y_B$ (top right) $Y_A$ (bottom left) and the baryon density $\rho_B$ (bottom right) as a function of $\mu_B$ at $c=0.0$ and $T=1.6$ for $L=32$.\label{fig5}}
}

\FIGURE[t]{
\begin{tabular}{cc}
\begin{minipage}[c]{0.45\textwidth}
\includegraphics[width=\textwidth]{fig6a}
\end{minipage}
&
\begin{minipage}[c]{0.45\textwidth}
\includegraphics[width=\textwidth]{fig6b}
\end{minipage}
\end{tabular}
\caption{
Plots of $Y_c$,$Y_B$ and $Y_A$ as a function of $L$ at $c=0$, $\mu_B=0.8$ and $T=1.6$ (left) and $T=1.9$ (right). \label{fig6}}
}

\TABLE[ht]{
\begin{tabular}{|c|c|c|c|c|c}
\hline
$\mu_B$ & 0.7 & 0.8 & 0.9 & 1.0 \\
$T_c$ & 1.793(1) & 1.7879(6) & 1.7338(4) & 1.6485(3) \\
$\chi^2/DOF$ & 4.39 & 1.66 & 1.39 & 1.15 \\
\hline
\end{tabular}
\caption{Results for $T_c$ obtained from fitting to the $XY$ scaling at $c=0$. \label{tb:O2_fitting_c0.0}}
}

\FIGURE[b]{
\begin{tabular}{cc}
\begin{minipage}[c]{0.45\textwidth}
\includegraphics[width=\textwidth]{fig7a}
\end{minipage}
&
\begin{minipage}[c]{0.45\textwidth}
\includegraphics[width=\textwidth]{fig7b}
\end{minipage}
\\
\begin{minipage}[c]{0.45\textwidth}
\includegraphics[width=\textwidth]{fig7c}
\end{minipage}
&
\begin{minipage}[c]{0.45\textwidth}
\includegraphics[width=\textwidth]{fig7d}
\end{minipage}
\end{tabular}
\caption{Plot of $LY_B$ (left) and $LY_A$ (right) at $\mu_B=0.8$ (top row) and $\mu_B=0.9$ (bottom row) as a function of $T$. The solid lines are joint fits to Eq.~\eqref{eq:scaling} with $\nu=0.6715$ fixed to the 3d $XY$ universal value.\label{fig7}}
}

In figure~\ref{fig6}, we plot the $L$ dependence of $Y_i$ below the transition temperature ($T=1.6$, left) and above the transition temperature ($T=1.9$, right) at a fairly large chemical potential ($\mu_B=0.8$). Clearly, $Y_B$ and $Y_A$ are non zero in the superfluidity phase, but zero in the symmetric phase. Since the relevant symmetry now is $U_A(1) \times U_B(1)$, the phase transition can be second order in the universality class of the $XY$ model \cite{Pelissetto:2000ek}. Assuming the transition is second order we fit the $Y_B$ and $Y_A$ data close to the transition to Eq.~\eqref{eq:scaling} at various values of $\mu_B \geq 0.7$. Unlike the $\mu_B=0$ case, now all the curves cross do cross at a point. We fix the critical exponent $\nu=0.6715$ which is the 3d $XY$ universal value \cite{Pelissetto:2000ek}. The value of $T_c$ at various $\mu_B$ along with the $\chi^2/DOF$ is given in table~\ref{tb:O2_fitting_c0.0}. The values of the other constants obtained in the fit for the $Y_B$ data at $\mu_B=0.8$ is $f_0 = 0.308(5)$, $f_1=-0.149(2)$, $f_2=0.0161(1)$, $f_3=0.0009(4)$ with $f_k=0, k \geq 4$. The quality of the fit can be observed in figure~\ref{fig7} where we plot $LY_B$ and $LY_A$ at $\mu_B=0.8$ and $\mu_B=0.9$.

At $\mu_B=0.7$ we find $\chi^2/DOF=4.39$ indicating the failure of the second order scaling. In figure~\ref{fig8} we plot $LY_B(L)$ as a function of $T$ for different volumes at $\mu_B=0.6$. We observe two transitions, one at $T \approx 1.58$ from the chirally broken phase to the superfluid phase, and the other at $T=1.75$ from the superfluid phase to a symmetric phase. A second order finite size scaling analysis again gives us a large $\chi^2/DOF$ at both the transitions indicating that both are first order. The existence of a large fluctuations in the $L=32$ data is due to tunneling between two metastable phases.

\FIGURE[t]{
\includegraphics[width=0.7\textwidth]{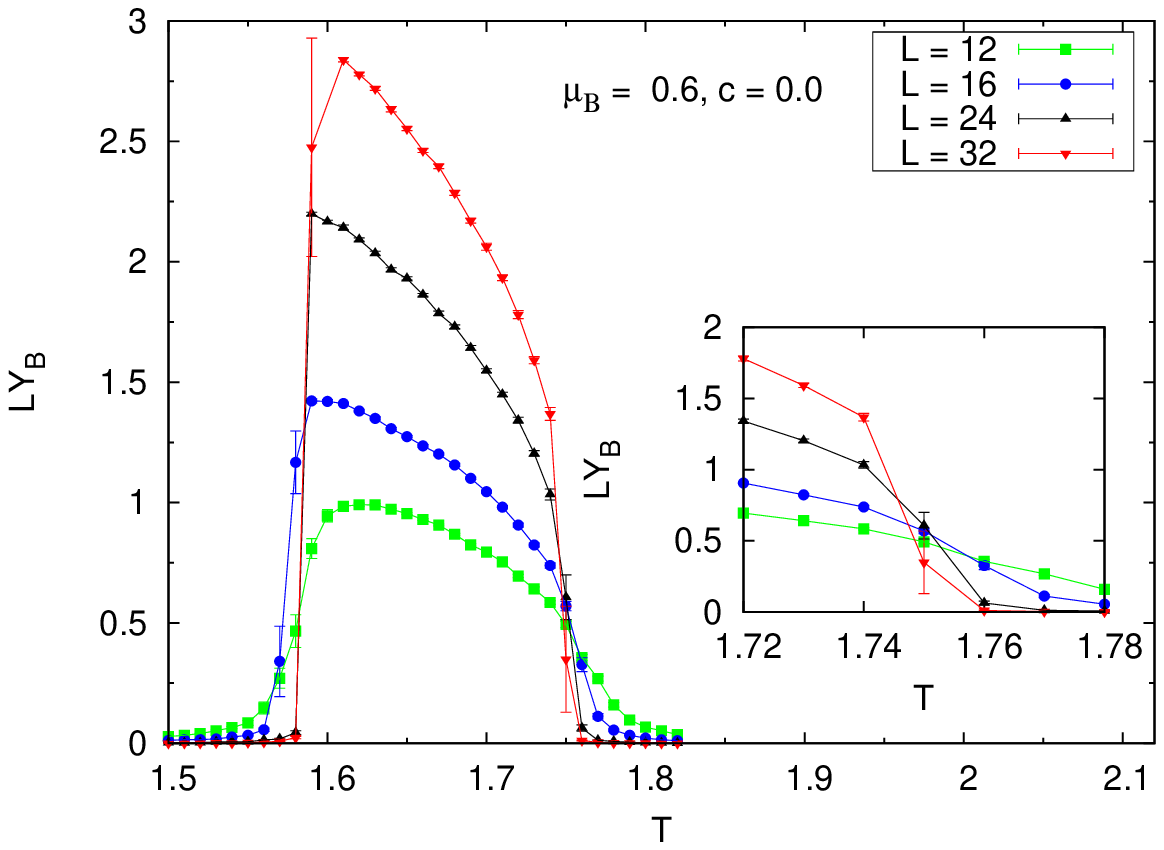}
\caption{Plot of  $LY_B$ as a function of $T$ at $\mu_B=0.6$. Two first transitions are seen: one from the chirally broken phase to the baryon superfluid phase at $T \sim 1.58$ and another from from superfluid phase to the high temperature symmetric phase at $T \sim 1.75$.\label{fig8}}
}

\FIGURE[t]{
\begin{tabular}{cc}
\begin{minipage}[c]{0.42\textwidth}
\includegraphics[width=\textwidth]{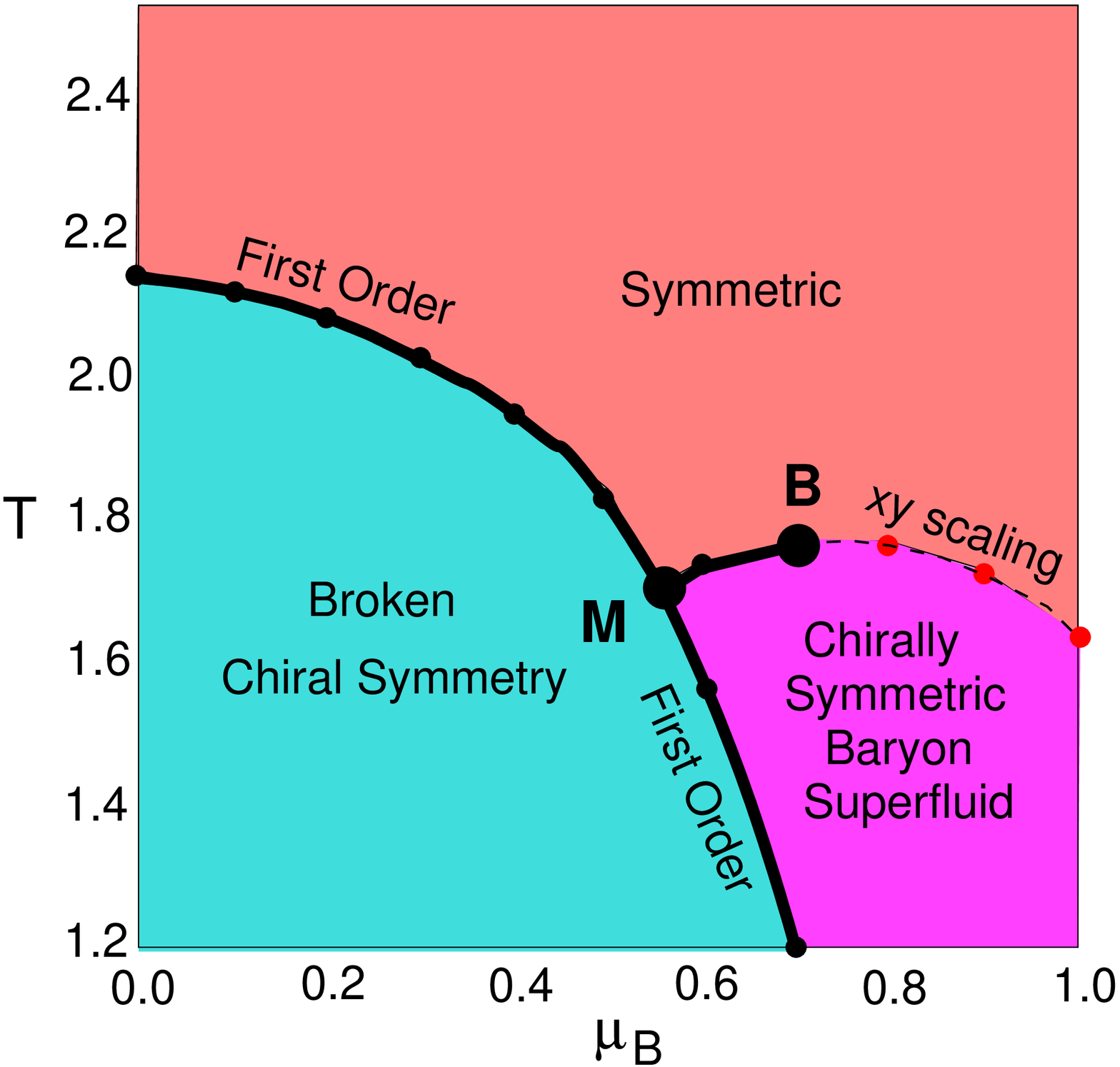}
\end{minipage}
&
\begin{minipage}[c]{0.45\textwidth}
\includegraphics[width=\textwidth]{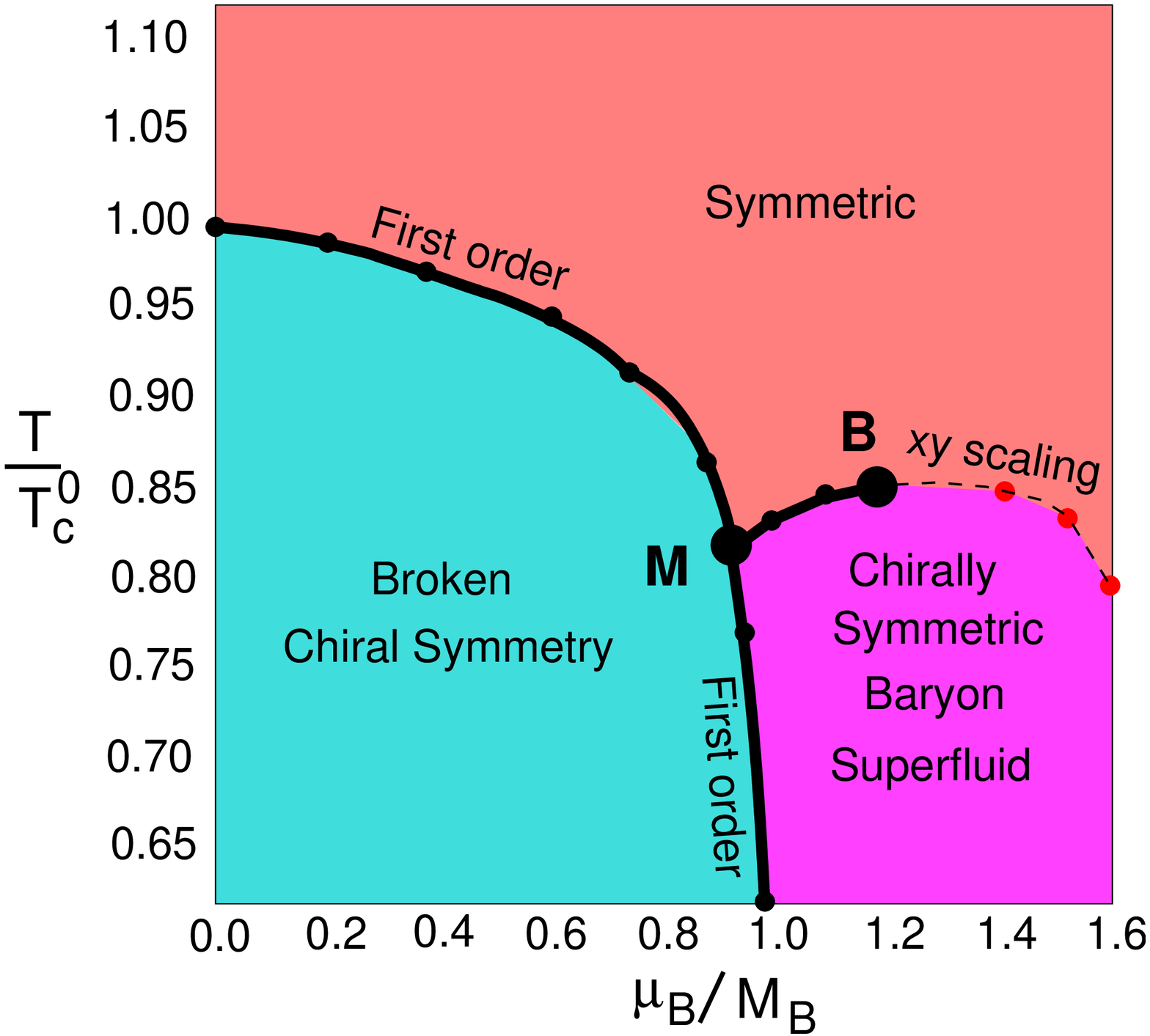}
\end{minipage}
\end{tabular}
\caption{Sketch of the phase diagram at $c=0$ as a function of $T$ and $\mu_B$ (left) and as a function of rescaled variables $T/T_c(\mu_B=0)$ and $\mu_B/M_B$ (right). Solid black lines first order lines and the dashed lines are second order lines. The point M is a three phase coexistence point and the point B is a tricritical point.\label{fig9}}
}

Using the above information, in figure~\ref{fig9} on the left we sketch the entire phase diagram as a function of  $T$ and $\mu_B$ in the absence of the anomaly. We find three phases, (1) A low $T$ and low $\mu_B$ phase where $SU(2) \times SU(2) \times U_A(1)$ is spontaneously broken into a flavor diagonal $SU(2)$ symmetry. This is the chirally broken phase, (2) a low $T$ large $\mu_B$ baryon superfluid superfluid phase driven by the spontaneous breaking of $U_B(1)$ symmetry. Since baryons carry axial charge, $U_A(1)$ symmetry also remains broken, and (3) a high temperature symmetric phase. For $\mu_B < 0.8$, all the phase transitions are first order. There is a point $M$ located roughly at $T = 1.70(5)$ and $\mu_B = 0.57(1)$ where all the three phases coexist. The first order finite temperature phase transition line from the superfluid phase to the symmetric phase ends at a tricritical point $B$ after which the transition line turns second order in the $XY$ universality class.

 Since the renormalized baryon mass changes with the paramter $T$ we think it would be interesting to plot the phase diagram in terms of the dimensionless variable $\mu_B/M_B(T)$. Further since $T$ is also not the temperature but simply a parameter in the model, it would be better to use $T/T_c^0$ where $T_c^0$ is the value of $T$ at the phase transition when $\mu_B=0$. Hence, in figure~\ref{fig9} on the right we again plot the phase diagram in these scaled variables. Remarkably, we now find that superfluidity sets in when $\mu_B/M_B \sim 1$ which means that binding energy between nucleons, if any, is small in our model.

\section{Phase Diagram with Anomaly}
\label{largeanomaly}

Next we set $c=0.3$ so that the $U_A(1)$ symmetry is explicitly broken by a large amount. Now, the action is only invariant under an $SU(2)\times SU(2)\times U_B(1)$ symmetry. At small $T$ and $\mu_B$ we expect the $SU(2)\times SU(2)$ symmetry to be spontaneously broken to the diagonal $SU(2)$ flavor group. As the temperature increases a phase transition from the chirally broken phase to the symmetric phase must occur at a critical temperature. If this phase transition is second order it would belong to the 3d $O(4)$ universality class~\cite{Chandrasekharan:2007up,Pelissetto:2000ek}. We have verified that our data fits well with this expectation for $0.0 \leq \mu_B \leq 0.5$ by fitting the $Y_c$ data to the scaling form given by Eq.~\eqref{eq:scaling}. We fix $\nu=0.745$, the $O(4)$ critical exponent, in the fits. The fitting results are tabulated in table~\ref{tb:O4_fitting_c0.3}. In figure~\ref{fig10} we plot the data along with the fits at $\mu_B=0.2$ and $\mu_B=0.5$ in order to show the quality of the fits.

\FIGURE[t]{
\begin{tabular}{cc}
\begin{minipage}[c]{0.45\textwidth}
\includegraphics[width=\textwidth]{fig10a}
\end{minipage}
&
\begin{minipage}[c]{0.45\textwidth}
\includegraphics[width=\textwidth]{fig10b}
\end{minipage}
\end{tabular}

\caption{Plot of $LY_C$ as a function of $T$ at $\mu_B=0.2$ (left) and $\mu_B=0.5$ (right). The solid lines are joint fits to the scaling form given in Eq.~\eqref{eq:scaling} with $\nu=0.745$ fixed to the 3d $O(4)$ universal value.\label{fig10}}
}

\DOUBLETABLE[ht]{
\begin{tabular}{|c|c|c|}
\hline
$\mu_B$ & $T_c$ & $\chi^2$/DOF \\ 
\hline
0.0 & 2.4753(4) & 0.75\\
0.1 & 2.4568(4) & 1.28\\
0.2 & 2.4078(6) & 1.09\\
0.3 & 2.3200(5) & 1.53\\
0.4 & 2.1958(4) & 1.20 \\
0.5 & 2.0351(5) & 1.67 \\
\hline
\end{tabular}}
{\begin{tabular}{|c|c|c|}
\hline
$\mu_B$ & $T_c$ & $\chi^2$/DOF \\ 
\hline
  0.6 & 2.0557(6) & 1.68\\
  0.7 & 2.0811(6) & 1.70\\
  0.8 & 2.0490(5) & 1.30\\
  0.9 & 1.9715(5) & 1.20\\
  1.0 & 1.8609(4) & 1.09\\
\hline
\end{tabular}}
{Fitting results of $Y_C$ for $O(4)$ scaling at $c=0.3$. \label{tb:O4_fitting_c0.3}}
{Fitting results of $Y_B$ for $XY$ scaling at $c=0.3$. \label{tb:O2_fitting_c0.3}}

Similar to the $c=0.0$ case, we again find that at low $T$ as $\mu_B$ increases there is a first order phase transition from the chirally broken phase to a chirally symmetric baryon superfluid phase where $U_B(1)$ is spontaneously broken. Again, the symmetry is restored at high temperatures through a second order phase transition which belongs to the 3d $XY$ universality class for sufficiently high $\mu_B$. If we repeat the scaling analysis of Eq.~(\ref{eq:scaling}) on our $Y_B$ data with $\nu=0.6715$ (the 3d XY critical exponent) we again find excellent fits for $\mu_B \geq 0.6$. The results are given in table~\ref{tb:O2_fitting_c0.3}. As an illustration of the quality of the fits we show the joint fit results at $\mu_B=0.7$ and $\mu_B=0.9$ in figure~\ref{fig11}.

\FIGURE[t]{
\begin{tabular}{cc}
\begin{minipage}[c]{0.45\textwidth}
\includegraphics[width=\textwidth]{fig11a}
\end{minipage}
&
\begin{minipage}[c]{0.45\textwidth}
\includegraphics[width=\textwidth]{fig11b}
\end{minipage}
\end{tabular}
\caption{Plots of $LY_B$ as a function of $T$ at $\mu_B=0.7$ (left) and $\mu_B=0.9$ (right). The solid lines are joint fits to the scaling form given in Eq.~\eqref{eq:scaling} with $\nu=0.6715$ fixed to the 3d $XY$ universal value.\label{fig11}}
}

Tables~\ref{tb:O4_fitting_c0.3} and \ref{tb:O2_fitting_c0.3} show the location of $T_c$ as a function of $\mu_B$ obtained from the $O(4)$ scaling and the $XY$ scaling respectively. What happens in the region $0.5 < \mu_B < 0.6$ is unclear. Unlike the results in the previous section we are unable to find a region where the two second order transitions clearly become first order. Can the two different second order transitions meet at a multicritical point? This seems unlikely, but cannot be ruled out with our limited statistics. Most likely the transition lines becomes weakly first order before meeting. Based on this guess, in figure~\ref{fig12} we sketch the phase diagram of the model at $c=0.3$. As compared to figure \ref{fig9}, the phase diagram in figure~\ref{fig12} contains the extra tricritical point $C$. Through a more extensive calculation, the precise nature and location of the points $M$, $B$ and $C$ can be understood.

\FIGURE[t]{
\begin{tabular}{cc}
\begin{minipage}[c]{0.42\textwidth}
\includegraphics[width=\textwidth]{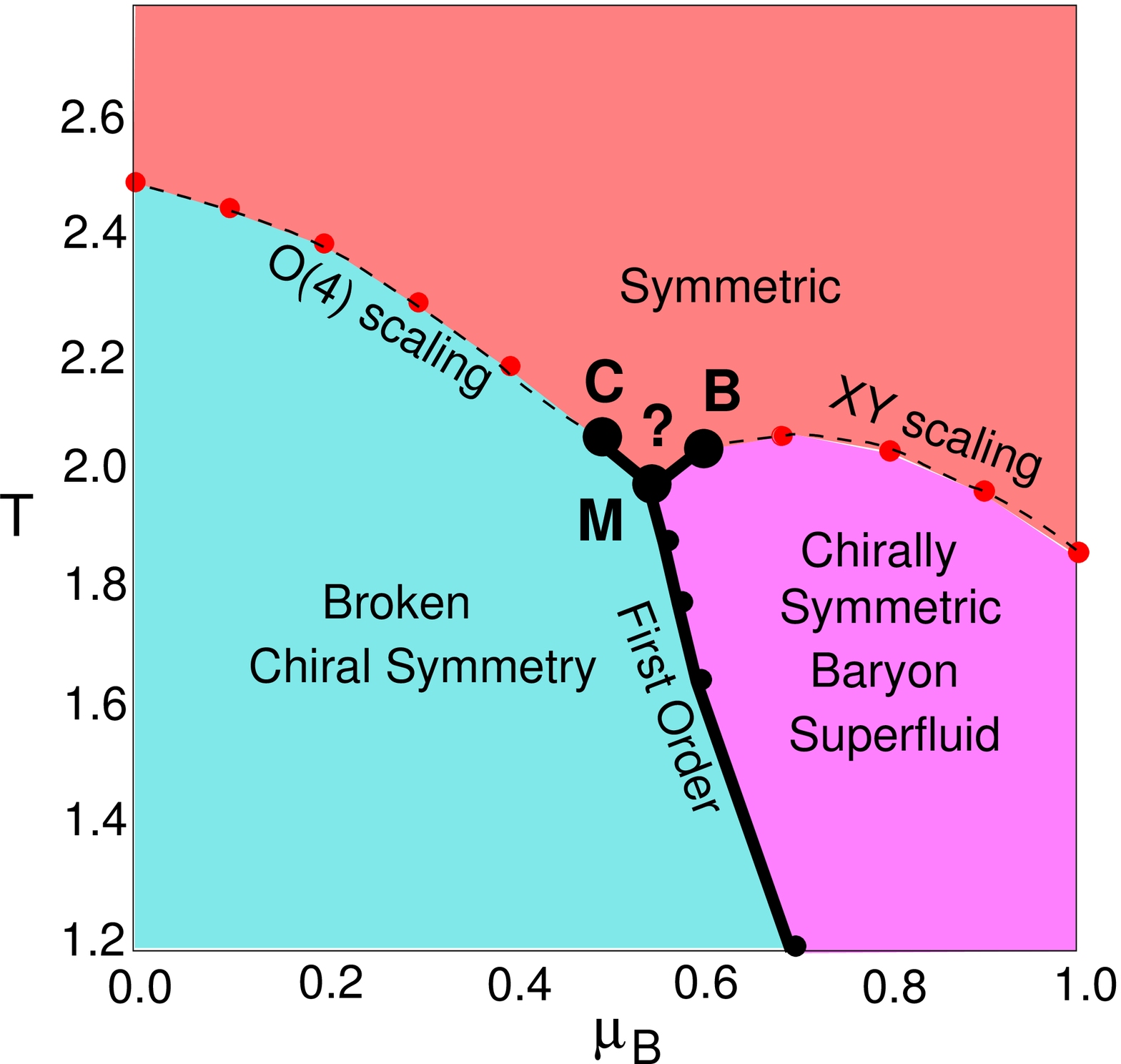}
\end{minipage}
&
\begin{minipage}[c]{0.45\textwidth}
\includegraphics[width=\textwidth]{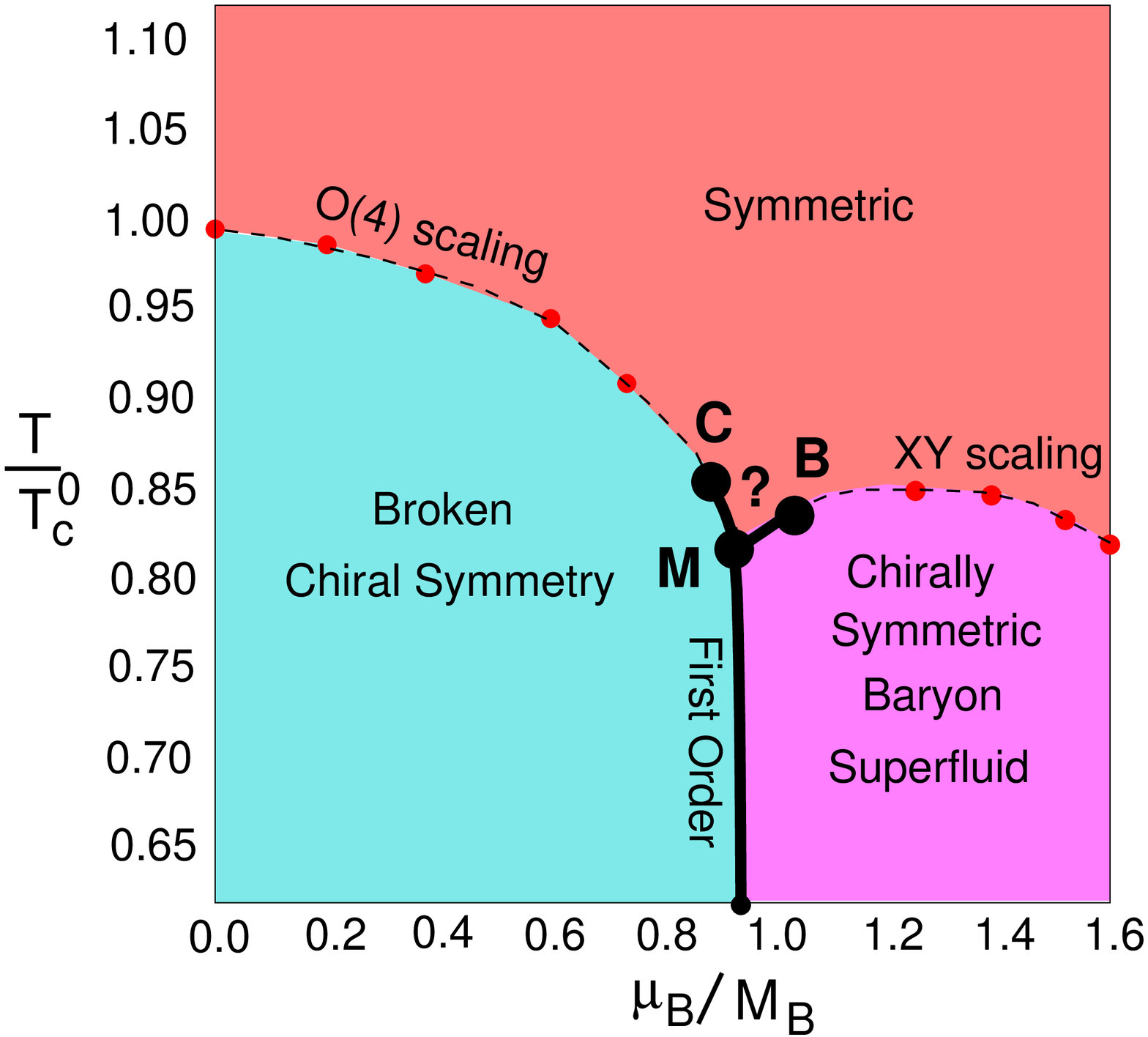}
\end{minipage}
\end{tabular}
\caption{Sketch of the phase diagram at $c=0.3$ as a function of $T$ and $\mu_B$ (left) and as a function of rescaled variables $T/T_c(\mu_B=0)$ and $\mu_B/M_B$ (right). Solid black lines represent first order lines while dashed line represent second order lines. The point M is a three phase coexistence point and the points C,B are tricritical points. The precise nature and location of $M$, $C$ and $B$ is a guess represented by a question mark.\label{fig12}}
}

Having established the phase diagrams for zero anomaly and a large anomaly, we next focus on how the phase diagram changes from figure \ref{fig9} in the absence of the anomaly to figure \ref{fig12} in the presence of a large anomaly. Note that two phase diagrams are very similar except for the nature of the finite temperature transition from the chirally broken phase to the symmetric phase. Without the anomaly the transition is first order, while in the presence of a large anomaly the transition becomes second order in the $O(4)$ universality class. So how does the transition change from first order to second order as the anomaly is slowly increased? Different scenarios have been proposed based on an effective theory approach \cite{Bowman:2008kc}. The conventional scenario is that the first order transition becomes weaker at $\mu_B = 0$ until it becomes second order above a critical value of the anomaly strength. Then, as the strength of the anomaly further increases the second order tricritical point $C$ moves into the $T-\mu_B$ plane. However, an alternative exotic scenario suggests that the first order transition becomes weak somewhere in the middle of the first order curve. As the anomaly increases further the second order line appears in the middle of the first order line creating two tricritical points (see figure 4, in \cite{Bowman:2008kc}). What happens in our model? Motivated by this question, we have studied the model at $c=0.005,0.01,0.02,0.03$ and $0.05$. We have evidence that in our model the change in the phase diagram follows the conventional scenario. To demonstrate this we discuss the results at $\mu_B=0.02$ below.

\FIGURE[b]{
\begin{tabular}{cc}
\begin{minipage}[c]{0.45\textwidth}
\includegraphics[width=\textwidth]{fig13a}
\end{minipage}
&
\begin{minipage}[c]{0.45\textwidth}
\includegraphics[width=\textwidth]{fig13b}
\end{minipage}
\end{tabular}
\caption{Plots of $LY_c$ as a function of $T$ at $\mu_B=0$ (left) and $\mu_B=0.1$ (right). The solid lines are joint fits to the $O(4)$ scaling form given in Eq.~\eqref{eq:scaling} with $\mu_B$ fixed to the value obtained from 3d $O(4)$ universality.\label{fig13}}
}

\FIGURE[t]{
\begin{tabular}{cc}
\begin{minipage}[c]{0.45\textwidth}
\includegraphics[width=\textwidth]{fig14a}
\end{minipage}
&
\begin{minipage}[c]{0.45\textwidth}
\includegraphics[width=\textwidth]{fig14b}
\end{minipage}
\end{tabular}
\caption{Plots of $LY_c$ as a function of $T$ at $\mu_B=0.3$ (left) and $\mu_B=0.5$ (right). The data does not fit well to the O(4) scaling form and is consistent with a first order transition.\label{fig14}}
}

\FIGURE[t]{
\includegraphics[width=0.6\textwidth]{fig15}
\caption{Monte Carlo time history of $Y_c$ data at $\mu_B=0.3$(left) and $\mu_B=0.5$ (right). The two state signal is clearly visible.\label{fig15}}
}

Figure \ref{fig13} shows that close to the finite temperature phase transition, $Y_c$ scales according to $O(4)$ universality at $\mu_B=0.0$ and $0.1$. But this scaling breaks down at $\mu_B=0.3$ and $0.5$ as can be seen in figure \ref{fig14}. In particular the large fluctuations in the $L=32$ data is consistent with the values fluctuating between the two metastable phases. Figure \ref{fig15} shows the time histories of the $Y_c$ data at $L=32$ for $\mu_B=0.3$ and $\mu_B=0.5$. These time histories clearly show the two state signal. Thus, we predict a tricritical point at roughly $\mu_B \sim 0.25(5)$. On the high chemical potential side, the anomaly has little effect on the phase diagram once it is plotted in the scaled variables (compare the right figures \ref{fig9} and \ref{fig13}). It does change the location of the point $B$ but only slightly. Thus, the phase diagram at $c=0.02$ at large $\mu_B$ should be very similar to the one at $c=0.0$. Our data confirms this.

\FIGURE[ht]{
\includegraphics[width=0.7\textwidth]{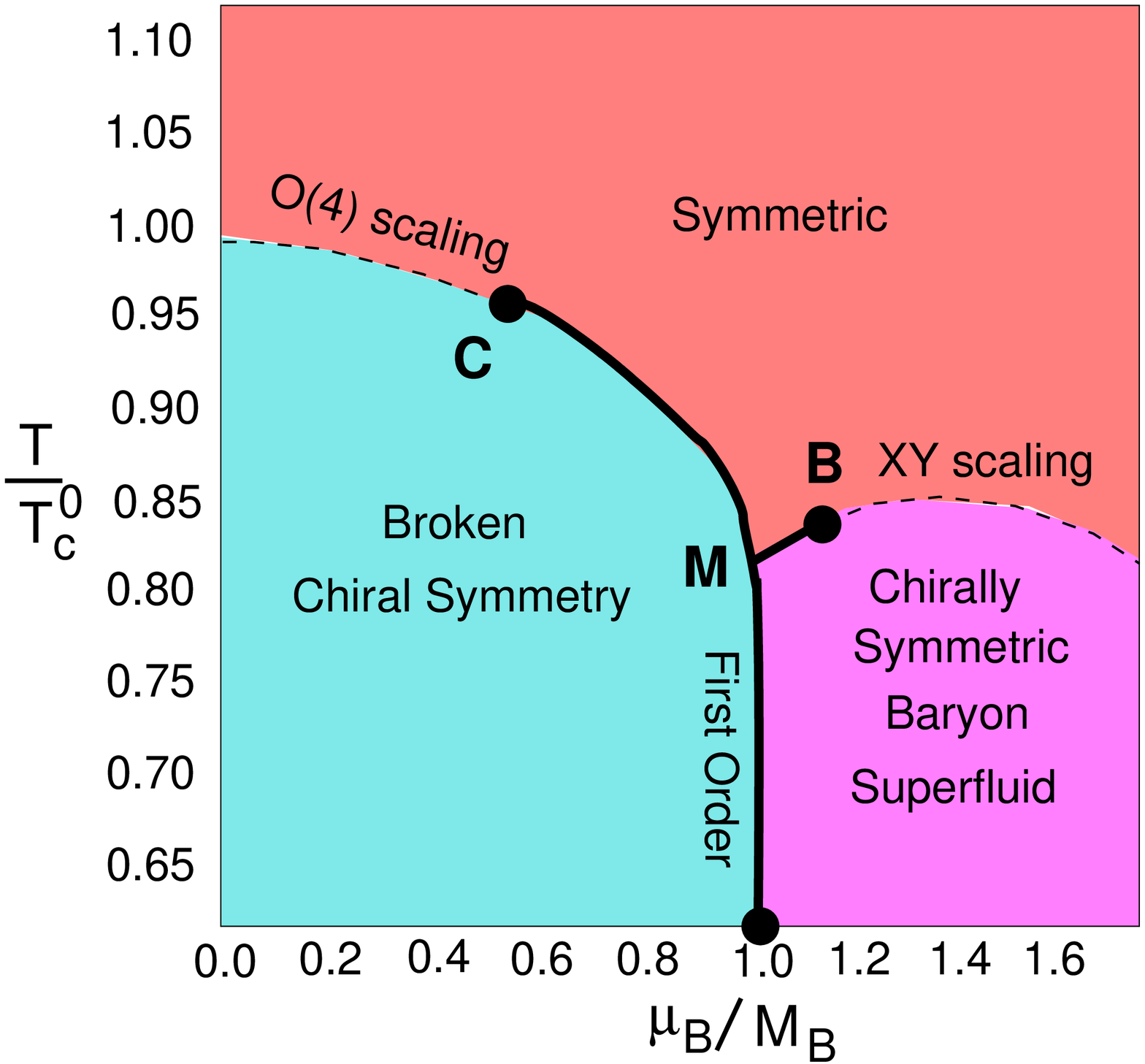}
\caption{Sketch of the phase diagram at a generic value of the anomaly strength. The first order critical lines are shown with solid critical lines are shown with a dashed line. The location of the critical point $C$ in the diagram is strongly dependent on the strength of the anomaly, while the location of the critical lines (both first and second order) and other critical points are insensitive to it.\label{fig16}}
}

\section{Conclusions}
\label{concl}

In this work we have computed the $T-\mu_B$ phase diagram of a lattice model which is invariant under an $SU(2)\times SU(2) \times U_A(1) \times U_B(1)$ symmetry and contains massive bosonic baryons. We also studied the effects of breaking the $U_A(1)$ symmetry. Due to the global symmetry and the nature of the baryons, our model may be considered as an interesting toy model of QCD with an even number of colors and may share some of the important phases and phase transitions with it. The term that breaks the $U_A(1)$ symmetry is mapped to the anomaly in QCD. Based on the results presented in this work, we conclude that the phase diagram contains at least three important phases : (1) a low temperature chirally broken phase with massive baryons (2) A chirally symmetric baryon superfluid phase at a moderately high chemical potential and low temperatures and (3) a symmetric high temperature phase. A qualitative sketch of the phase diagram at a generic value of the anomaly strength is shown in figure \ref{fig16}. We find that the location of the tricritical point $C$ in the figure is strongly dependent on the strength of the anomaly. There is also a tricritical point on the baryon superfluid side (point $B$) and a point $M$ where all the three phases coexist. If the points $B$ and $M$ exist in real QCD the physics close to them could also be very interesting both theoretically and phenomenologically.

In the presence of a quark mass the finite temperature $O(4)$ transition will turn into a cross over, while the tricritical point $C$ will turn into a critical end point. The location of a similar critical end point in QCD is the main focus of many calculations today including one of the goals of the experimental program at RHIC \cite{Aggarwal:2010cw}. Among the many dynamical effects that affect its location, here we find that the strength of the anomaly is an important one. Given that the anomaly in lattice QCD calculations is strongly dependent on the lattice spacing, we believe it will be rather difficult to find the precise location of the critical point in real QCD without controlling all the systematic errors. On the other hand, the two other critical points $B$ and $M$ seem rather robust and may be easier to locate.

The present work can be extended further in many directions. We believe that the simplicity of the phase diagram in our model comes from the fact that baryons in our model were singlets under chiral transformations. One can build and study lattice field theory models where baryons transform non-trivially under the chiral symmetry. It would be interesting to see if these models contain a phase where both chiral symmetry and the baryon number symmetry are both spontaneously broken. Such a phase would make the phase diagram even more interesting. One can also compute the properties of nuclei and their physics within our toy model. It would also be interesting to understand the connection between the strong first order transition as a function of the chemical potential and the binding of the baryons. Finally we wish to emphasize that new methods of analysis to locate critical points in QCD, can first be tested in toy models such as the one studied here before being applied to QCD.

\acknowledgments

S.C. would like to thank, Michael Fromm, Sourendu Gupta and Philippe de Forcrand for helpful conversations. This work was supported in part by the Department of Energy grant DE-FG02-05ER41368. The computations were performed on the local cluster funded by the Department of Energy.

\appendix

\section{Comparison between exact results and the algorithm}

In order to confirm that the algorithm reproduces the results in the model accurately, we have obtained analytic expressions for the partition function, the axial current susceptibility ($Y_A$), the baryon current susceptibility ($Y_B$), and the chiral current susceptibility ($Y_C$) an a $2\times 2$ lattice. These are given below:
\begin{eqnarray}
Z &=& 4 (\omega_D)^2(1+T^4) 
+ 8 \omega_D (2\{1 + T^4\} + \mathrm{e}^{-2 m_B}\{1 + T^4\cosh(2\mu_B)\}) 
+ 16(1+4T^2+T^4)
\nonumber \\
&& + 16\mathrm{e}^{-2m_B}\{1+T^4\cosh(2\mu_B)\} 
 + 4\mathrm{e}^{-4m_B}\{1 + 4[1+T^2\cosh(2\mu_B)] + T^4\cosh^2(2\mu_B)\}
\nonumber \\
&& 
+ 4 c^2 \omega_D (1+T^2) 
+ 4 c^2 (2 + 2 T^2 + \mathrm{e}^{-2m_B}\{1 + T^2\cosh(2\mu_B)\}) + c^4
\end{eqnarray}

\begin{eqnarray}
Z\times Y_A &=& 8 (\omega_D)^2 + 8 \omega_D (2+\mathrm{e}^{-2m_B}) + 8T^2(2+\mathrm{e}^{-4m_B}) + 8T^2[2+\mathrm{e}^{-4m_B}\cosh(2\mu_B)]  
\nonumber \\ 
&& + 4c^2\omega_D \\
\nonumber \\
Z\times Y_B &=& 4\mathrm{e}^{-2m_B}(2\omega_D + 4 + c^2) + 8\mathrm{e}^{-4m_B}[1 + T^2 + T^2\cosh(2\mu_B)]\\
\nonumber \\
Z\times Y_C &=& 16 + 8\mathrm{e}^{-2m_B} + 8\omega_D + 16T^2 + 4c^2;
\end{eqnarray}
In tables~\ref{tab4} and \ref{tab5}, we compare the analytic results with the results obtained using our algorithm for different parameters. The agreement gives us confidence that our algorithm must be correct.

\TABLE[]{
\begin{tabular}{cccccccc}
\hline
\hline
$m_B$ & $\mu_B$ &\textit{Exact} & \textit{Algorithm} &\textit{Exact} & \textit{Algorithm}&\textit{Exact} & \textit{Algorithm} \\
\hline
 & & \multicolumn{2}{c}{$Y_A$}  &  \multicolumn{2}{c}{$Y_B$} &  \multicolumn{2}{c}{$Y_C$}\\
\hline
\multicolumn{8}{c}{$T=1.0,\ c=0.0,\ \omega_D = 0.0$}\\
\hline
0.0 & 0.0 & 0.28571429 & 0.28573(2) & 0.23809524 & 0.23808(2) & 0.23809524 & 0.23808(2) \\
0.0 & 0.5 & 0.27419502 & 0.27420(2) & 0.23228892 & 0.23228(2) & 0.20953053 & 0.20951(2) \\
0.0 & 1.0 & 0.22684802 & 0.22684(2) & 0.20095862 & 0.20096(2) & 0.12944702 & 0.12947(1) \\
0.2 & 0.0 & 0.28938325 & 0.28939(2) & 0.15882791 & 0.15883(1) & 0.27589442 & 0.27589(2) \\
0.2 & 0.5 & 0.27867066 & 0.27869(2) & 0.15891388 & 0.15892(1) & 0.25307451 & 0.25305(2) \\
0.2 & 1.0 & 0.23552474 & 0.23552(2) & 0.15074712 & 0.15076(1) & 0.17915523 & 0.17915(2) \\
\hline
\multicolumn{8}{c}{$T=1.0,\ c=0.0,\ \omega_D = 1.0$}\\
\hline
0.0 & 0.0 & 0.35714286 & 0.35715(2) & 0.21428571 & 0.21427(1) & 0.21428571 & 0.21428(1) \\
0.0 & 0.5 & 0.33570326 & 0.33569(2) & 0.20833887 & 0.20833(1) & 0.19104659 & 0.19105(1) \\
0.0 & 1.0 & 0.26374692 & 0.26376(2) & 0.18108186 & 0.18109(1) & 0.12399759 & 0.12400(1) \\
0.2 & 0.0 & 0.36826413 & 0.36824(2) & 0.14435557 & 0.14434(1) & 0.24369015 & 0.24366(1) \\
0.2 & 0.5 & 0.35029205 & 0.35030(2) & 0.14320792 & 0.14323(1) & 0.22537934 & 0.22539(1) \\
0.2 & 1.0 & 0.28633667 & 0.28636(2) & 0.13426676 & 0.13426(1) & 0.16550478 & 0.16549(1) \\
\hline
\multicolumn{8}{c}{$T=1.0,\ c=0.2,\ \omega_D = 0.0$}\\
\hline
0.0 & 0.0 & 0.28408822 & 0.28413(2) & 0.23768714 & 0.23769(2) & 0.23768714 & 0.23767(2) \\
0.0 & 0.5 & 0.27269729 & 0.27272(2) & 0.23185364 & 0.23186(2) & 0.20921955 & 0.20922(2) \\
0.0 & 1.0 & 0.22582231 & 0.22582(1) & 0.20056541 & 0.20057(2) & 0.12937716 & 0.12939(1) \\
0.2 & 0.0 & 0.28756535 & 0.28756(2) & 0.15861716 & 0.15861(1) & 0.27533532 & 0.27534(2) \\
0.2 & 0.5 & 0.27695539 & 0.27695(2) & 0.15865772 & 0.15866(1) & 0.25259387 & 0.25258(2) \\
0.2 & 1.0 & 0.23423048 & 0.23423(2) & 0.15043018 & 0.15042(1) & 0.17893372 & 0.17892(2) \\
\hline
\multicolumn{8}{c}{$T=1.0,\ c=0.2,\ \omega_D = 1.0$}\\
\hline
0.0 & 0.0 & 0.35582134 & 0.35581(2) & 0.21377689 & 0.21376(1) & 0.21377689 & 0.21376(1) \\
0.0 & 0.5 & 0.33451804 & 0.33453(2) & 0.20784361 & 0.20784(1) & 0.19064501 & 0.19065(1) \\
0.0 & 1.0 & 0.26298931 & 0.26298(2) & 0.18069067 & 0.18070(1) & 0.12385944 & 0.12386(1) \\
0.2 & 0.0 & 0.36680615 & 0.36682(2) & 0.14402179 & 0.14402(1) & 0.24301429 & 0.24302(1) \\
0.2 & 0.5 & 0.34894700 & 0.34894(2) & 0.14286463 & 0.14286(1) & 0.22479567 & 0.22482(1) \\
0.2 & 1.0 & 0.28538737 & 0.28536(2) & 0.13393857 & 0.13393(1) & 0.16520110 & 0.16519(1) \\
\hline
\end{tabular}
\caption{Comparison of exact results with results from the algorithm at $T=1.0$.\label{tab4}}
}

\newpage

\TABLE[]{
\begin{tabular}{cccccccc}
\hline
\hline
$m_B$ & $\mu_B$ &\textit{Exact} & \textit{Algorithm} &\textit{Exact} & \textit{Algorithm}&\textit{Exact} & \textit{Algorithm} \\
\hline
 & & \multicolumn{2}{c}{$Y_A$}  &  \multicolumn{2}{c}{$Y_B$} &  \multicolumn{2}{c}{$Y_C$}\\
\hline
\multicolumn{8}{c}{$T=1.5,\ c=0.0,\ \omega_D = 0.0$}\\
\hline
0.0 & 0.0 & 0.24870466 & 0.24870(2) & 0.13816926 & 0.13816(1) & 0.13816926 & 0.13816(1) \\
0.0 & 0.5 & 0.22401091 & 0.22400(2) & 0.13271409 & 0.13272(1) & 0.11412102 & 0.11413(1) \\
0.0 & 1.0 & 0.15405340 & 0.15406(1) & 0.10716914 & 0.10715(1) & 0.05860532 & 0.05859(1) \\
0.2 & 0.0 & 0.25245829 & 0.25246(2) & 0.08731264 & 0.08730(1) & 0.16423607 & 0.16422(1) \\
0.2 & 0.5 & 0.23135853 & 0.23135(2) & 0.08719658 & 0.08719(1) & 0.14336794 & 0.14337(1) \\
0.2 & 1.0 & 0.16653699 & 0.16655(2) & 0.07961861 & 0.07962(1) & 0.08643966 & 0.08644(1) \\
\hline
\multicolumn{8}{c}{$T=1.5,\ c=0.0,\ \omega_D = 1.0$}\\
\hline
0.0 & 0.0 & 0.23178808 & 0.23177(2) & 0.11258278 & 0.11258(1) & 0.11258278 & 0.11258(1) \\
0.0 & 0.5 & 0.20874557 & 0.20875(1) & 0.10839747 & 0.10839(1) & 0.09477320 & 0.09477(1) \\
0.0 & 1.0 & 0.14533254 & 0.14534(1) & 0.09017774 & 0.09019(1) & 0.05209065 & 0.05209(1) \\
0.2 & 0.0 & 0.23366106 & 0.23366(2) & 0.07128440 & 0.07128(1) & 0.12993783 & 0.12994(1) \\
0.2 & 0.5 & 0.21443609 & 0.21447(2) & 0.07078745 & 0.07081(1) & 0.11495121 & 0.11498(1) \\
0.2 & 1.0 & 0.15675114 & 0.15675(2) & 0.06521854 & 0.06523(1) & 0.07324665 & 0.07325(1) \\
\hline
\multicolumn{8}{c}{$T=1.5,\ c=0.2,\ \omega_D = 0.0$}\\
\hline
0.0 & 0.0 & 0.24781350 & 0.24784(2) & 0.13804130 & 0.13804(1) & 0.13804130 & 0.13804(1) \\
0.0 & 0.5 & 0.22326474 & 0.22326(2) & 0.13257534 & 0.13258(1) & 0.11404420 & 0.11404(1) \\
0.0 & 1.0 & 0.15366975 & 0.15366(1) & 0.10705814 & 0.10704(1) & 0.05861527 & 0.05862(1) \\
0.2 & 0.0 & 0.25145743 & 0.25147(2) & 0.08727235 & 0.08727(1) & 0.16404125 & 0.16404(1) \\
0.2 & 0.5 & 0.23048223 & 0.23050(2) & 0.08713335 & 0.08713(1) & 0.14322330 & 0.14322(1) \\
0.2 & 1.0 & 0.16602244 & 0.16601(1) & 0.07953373 & 0.07952(1) & 0.08641295 & 0.08640(1) \\
\hline
\multicolumn{8}{c}{$T=1.5,\ c=0.2,\ \omega_D = 1.0$}\\
\hline
0.0 & 0.0 & 0.23125599 & 0.23124(1) & 0.11246010 & 0.11245(1) & 0.11246010 & 0.11246(1) \\
0.0 & 0.5 & 0.20830747 & 0.20832(1) & 0.10827684 & 0.10828(1) & 0.09469566 & 0.09470(1) \\
0.0 & 1.0 & 0.14511317 & 0.14513(1) & 0.09008803 & 0.09009(1) & 0.05209047 & 0.05209(1) \\
0.2 & 0.0 & 0.23309400 & 0.23310(1) & 0.07122714 & 0.07123(1) & 0.12976315 & 0.12976(1) \\
0.2 & 0.5 & 0.21394944 & 0.21395(1) & 0.07072219 & 0.07072(1) & 0.11482041 & 0.11482(1) \\
0.2 & 1.0 & 0.15647860 & 0.15648(1) & 0.06515061 & 0.06515(1) & 0.07321454 & 0.07321(1) \\
\hline
\end{tabular}
\caption{Comparison of exact results with results from the algorithm at $T=1.5$.\label{tab5}}
}

\clearpage

\bibliography{myref}

\bibliographystyle{JHEP}

\end{document}